# LISTEN! A SMARTPHONE INQUIRY ON THE DOMINO EFFECT

**Authors**[1]: Laurent Dalla Pola[a], Luis Darmendrail[a,b], Edward Galantay[a], Andreas Müller[a,b]

**Author affiliations:**
[a] Department of Physics, University of Geneva, Switzerland
[b] Institute of Teacher Education, University of Geneva, Switzerland

## Abstract

In this work, we investigate the phenomenon of a chain of falling dominoes ("domino effect") by an acoustical measurement with a smartphone. Specifically, we will present an approximate model valid for a considerable range of realistic parameters of the domino effect; a new, easy and low-cost experimental method using acoustical data captured by smartphones; and a comparison of these experimental data to the simplified treatment and to other, more comprehensive theories.

The results of the experiments show good agreement with other measurements and theory, both advanced ones from the literature, and our own simplified treatment: qualitatively the approach to a constant propagation velocity of the domino front after a small number of toppled dominoes as predicted by theory; quantitatively the numerical value of the asymptotic propagation velocity.

The example can be seen as a demonstration of the use of mobile devices as experimental tools for student projects about a phenomenon of "everyday physics", and in the same time of interest for current research (the domino effect as a paradigm of collective dynamics with applications e.g. in neuronal signal transduction.

## 1 Introduction

In this work, we present the phenomenon of a chain of falling dominoes ("domino effect"), investigated by acoustical means with a smartphone. In general, the domino effect consists of systematically placing a series of dominoes in a row (see Fig. 1), sometimes forming artificial figures and mechanisms ("Rube Goldberg" like devices)[1,2,3], and by giving the initial domino a slight push, causing a chain effect with a propagating wave of toppling dominoes. The domino effect has been met with some interest as an example of "everyday physics" (ref. 4, expl. 1.68) and as a model system for collective dynamics[5,6,8].

As an alternative to the use of more complex apparatus (discussed below), acoustical measurements are taken with a smartphone to find the time of collisions between the leading domino piece and the next one, and to deduce the propagation speed from it. Results are compared with models published the literature, and a simplified one derived below (sect. 2.2).

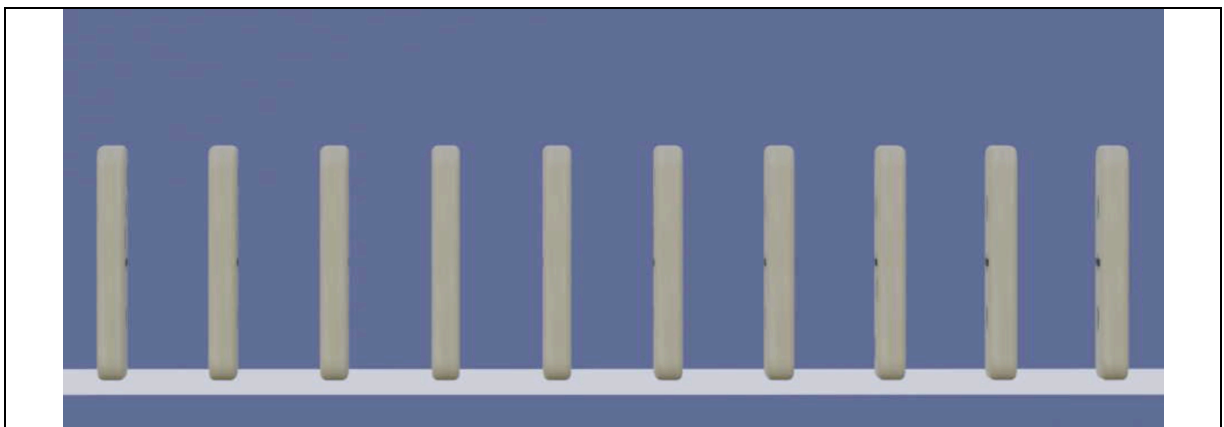

Fig. 1: Chain of dominoes

---

[1] Author contributions (according to CRediT, https://authorservices.wiley.com/author-resources/Journal-Authors/open-access/credit.html): LDP and EG have conceived the experiment and carried out the measurements; LD and AM have carried out the comparison to theory, with contributions by LDP and EG; LD and AM have prepared the manuscript, with contributions by LDP and EG.

## 2 Theory

We first give a brief account of previous treatments and then present an approximate model allowing for a simplified estimation of the relevant parameters.

### 2.1 Propagation velocity and toppling time

Once fully developed, the domino front propagates with a constant asymptotic velocity.[6,7] This velocity is given

(1) $\quad v_{as} = l / t_t$

where $l$ is the spacing between dominoes (see Fig. 2a), and $t_t$ is the (asymptotic) "toppling" time of the leading domino until it hits the next one (once the state of constant propagation speed is reached). Instead of $v$, the dimensionless velocity $\hat{v} = v / \sqrt{gh}$ is also often used, where $h$ is the height of the dominoes, and $g$ the gravitational acceleration.[7]

There are two main difficulties of an accurate determination of $t_t$ : First, anharmonicity: the toppling of a rod usually takes place over a large angle range of 90 degrees (from upright to horizontal) and is described by an anharmonic pendulum, requiring special functions (elliptic integrals) for the solution (9, ch. 21). Second, collective dynamics: the "toppling" is not that of an isolated domino, but that of the group of foremost dominoes, pushing the leading one.[7] An additional complication occurs if one does not assume "infinitely" thin dominos (i.e. disregarding all powers of $d/h$), leading e.g. to changes for the moment of inertia, a "potential barrier" in the toppling motion, etc.[7]

Depending of various assumptions on geometry, collision dynamics, and additional effects like friction and sliding, a number on theoretical and numerical[7,10,11,5,12,13,8] have been published, almost exclusively on very high levels of sophistication. Banks[9] has provided a model fully taking into account anharmonicity (but not collective effects). We note incidentally that Banks erroneously uses $v_{as} = (l-d) / t_t$ for the propagation speed (ref. 9, p267), involving the interspacing $s = l-d$ instead of the spacing $l$. Indeed, the leading domino only covers the distance $s$ in the time $t_t$, but the next domino it hits then immediately starts to topple around the pivot point at its *leading* edge (right edge in Fig. 2a), i.e. with an additional displacement of the size of its thickness $d$. Of course, the small delay for the elastic wave propagation through the dominoes is disregarded here (it is of the order of $d/v_{el.\ wave}$ ~ (some *mm* / some hundred *m/s*) ~ 10 μs, while the $t_t$ ~ 10 ms, see sect. 0). This leads to correction factors $1+d/s$ for $v_{as}$ and better consistency with experiment of the approach of Banks[9] than thought previously[7].

Van Leeuwen[7] has published a very complete theory taking into account collective effects, friction, and making no assumptions about the relative dimensions of the system (in particular covering finite thickness of the dominoes and also anharmonicity effects). In that theory, an important assumption made is that the leading domino, after hitting the following one, keeps pushing on it and does not bounce back. This is consistent with observation, and it means that the collisions are inelastic.

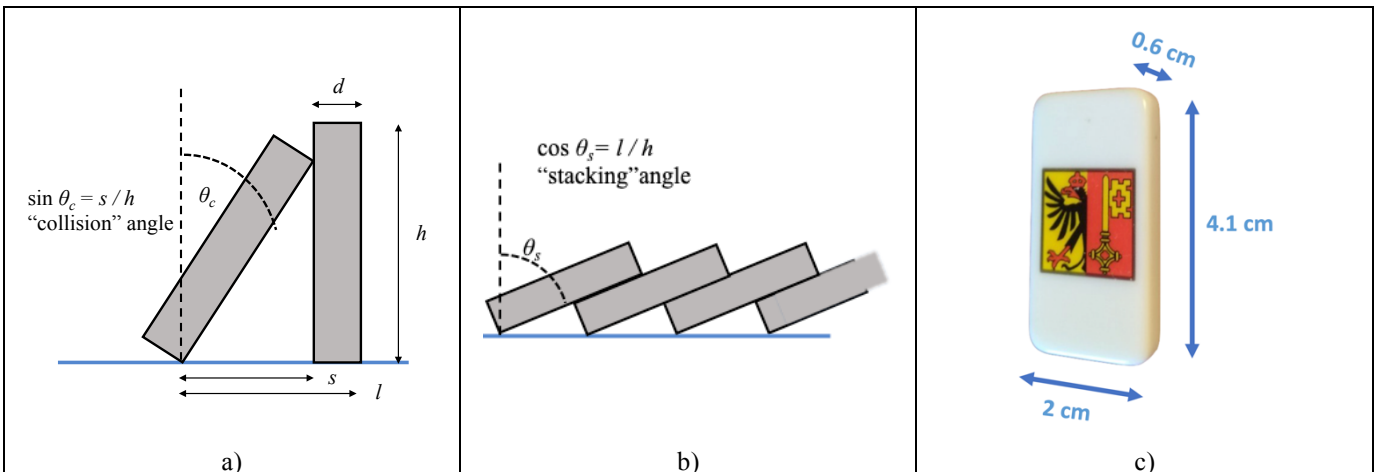

Fig. 2: Definition sketches (dimensions and geometry) of the domino effect
a) Two dominoes at the moment of collision; $h$, $l=s+d$, $s$, $d$ are the height, spacing, interspacing (gap width), and thickness of the dominoes; $\theta$ is the leaning angle, measured from the vertical; $\theta_c$ specifically that at the moment of the collision when the leading domino hits the next one.
b) Dominos in the final, "stacked" position.
c) Domino piece used in the experiments.

Moreover, this also takes into account the fact of the domino wave approaching a finite asymptotic velocity, while elastic collisions would lead to an ever increasing velocity as more and more energy is in the system by more and more dominoes having fallen and released their potential energy. Moreover, a couple of further simplifying assumptions are made on the shape, arrangement and motion of the dominoes, which

- all have the same shape, dimensions, and distance
- are in a straight line and the end position is exactly stacked (no sidewise deviation)
- do not slip on the surface, nor "take off" during the collision

When friction can be neglected, the theory of van Leeuwen (ref. 7, eqs. 47, 48) allows for the following closed form of the final result

$$(2) \quad v_{as} = \sqrt{gh} Q(d,h,s) K(d,h,s)$$
$$\approx \sqrt{gh} Q(d,h,s)$$

where

$$(3) \quad Q = \sqrt{\frac{3}{1+d^2/h^2}} \frac{s+d}{h} \frac{\sqrt{P(d,h,s)}}{\arcsin(s/h)}$$
$$P = \frac{sh - d\sqrt{s^2 + 2sd}}{h(s+d)}$$

In eq. (2), the factor $K$ is a correction factor close to 1 for usual working conditions of the domino effect (up to a 10% increase for separations as large as $s/h \geq 0.9$, which will be not considered here). Note that the scale (and dimension) for $v_{as}$ in eq. (2) is given by the factor $(gh)^{1/2}$, while $Q(d, h, s)$ is unitless. Below, we will compare results of our smartphone experiment and of a simplified theory (2.2) to van Leeuwen's predictions.

## 2.2 Approximate model of the toppling process

In the following, we will present an approximate treatment valid for a considerable range of realistic parameters of the domino effect. We start from the same simplifying assumptions as Banks[9]: (i) the propagation of the domino wave is considered as a sequence of single-domino events, where the next-but leading domino topples until it hits the leading one, which then topples independently of the others dominoes; (ii) linear momentum during the collisions is conserved. Based on this, and following from momentum conservation, the initial angular velocity is given by $\dot{\theta}_0 = \cos(\theta_c)\dot{\theta}_c$, where the subscripts (0 and $c$) refer to the moments of the beginning of the toppling and of the collision with the next domino, respectively (ref. 8, eq. 21.24).

We now consider the equation of motion of the leading domino. The toppling of a rod (the domino) is described by rotating freely (without sliding) around its basis, where the motion starts in the upright (or slightly leaning) position and ends in the horizontal condition, when the rod hits the ground. This leads to the equation of motion[14]

$$(4) \quad I\ddot{\theta} = gMR_S \sin\theta \text{ or}$$
$$\ddot{\theta} = \omega^2 \sin\theta, \text{ with } \omega = \sqrt{gMR_S/I}$$

($g$ = gravitational acceleration, and $M$, $R_S$, $I$ being the mass, centre-of-mass position and moment of inertia of the rod, respectively). For a thin rod of height $h$ one has $R_S = h/2$, $I = \frac{1}{3} Mh^2$, and $\omega = (\frac{2}{3} g/h)^{1/2}$; we will use the latter expression for the following. Note the "+" sign on the RHS of eq. (2) in contrast to the "–" sign of the ordinary pendulum: for toppling, the initial position is the (unstable) upright position, where gravity leads to an ever increasing inclination, whereas for the pendulum, gravity acts as restoring force to the equilibrium point with zero inclination.

The anharmonicity in the toppling of the domino as mentioned above occurs due to the wide range of $\theta$ from 0 (upright) to $\pi/2$ (horizontal) covered in a complete toppling process, necessarily involving higher orders in the $\sin\theta$ term in the accelerating force; this, in turn, involves elliptic integrals for the solution function[14]. The dynamics of the toppling process has several interesting applications and extensions having led to a whole series of publications, like the faster-than-gravity acceleration of tip of the toppling object[14,15,16], the breaking of falling chimneys at a specific position[17,18], or the inverted pendulum[19,20], in turn basis of many further applications like the stability of the upright position of Humans[21] or of the unicycle (ref. 4, expl. 1.116).

For the domino effect, the toppling motion of interest is that till collision with the next domino (see (see Fig. 2a), *not* till hitting the ground. This allows for an important simplification: in many settings (such as one studied experimentally in sect. 3), the angle $\theta_c$ at the moment of the collision with the leading domino is small enough to disregard higher powers in the sin term of the force law, i.e. anharmonicity. The equation of motion then becomes

(5) $\quad \ddot{\theta} = \omega^2 \theta$

with the general solution $\theta(t) = a \sinh(\omega t) + b \cosh(\omega t)$. For the toppling to begin, the rod must be either initially inclined ($\theta(0) \neq 0$), or set into motion ($\dot{\theta}(0) = \dot{\theta}_0 \neq 0$), or both. In the present case, the initial conditions are $\theta(0) = 0$, and $\dot{\theta}_0 = \cos(\theta_c)\dot{\theta}_c$, as given above. This leads to $a = \dot{\theta}_0/\omega$ and $b = 0$, yielding

(6) $\quad \theta(t) = (\dot{\theta}_0/\omega)\sinh(\omega t)$.

From this, we easily obtain the (approximate) result of the domino toppling time sought for by setting $\theta(t) = \theta_c$, and solving for $t$,

(7) $\quad t_t = \omega^{-1}\mathrm{arcsinh}(\frac{\omega}{\dot{\theta}_0}\theta_c)$.

For small $\theta_c$ (i.e. small spacings), a further simplification becomes possible. On the one hand, we have $\dot{\theta}_0 = \cos(\theta_c)\dot{\theta}_c$ from the initial condition (see above), on the other hand $\dot{\theta}(t) = \dot{\theta}_0 \cosh(\omega t)$ from (6), and thus $\dot{\theta}_c = \dot{\theta}(t_c) = \dot{\theta}_0 \cosh(\omega t_c)$. Combining the two equations, one obtains $\dot{\theta}_c = \cos(\theta_c)\dot{\theta}_c \cosh(\omega t_t)$ and then $\cos(\theta_c)\cosh(\omega t_t) = 1$. Expanding the arguments of the cos and cosh functions to lowest (quadratic) order, one has $(1-\frac{1}{2}\theta_c^2)(1+\frac{1}{2}(\omega t_c)^2) = 1$ and from this finally

(8) $\quad t_t \approx \omega^{-1}\theta_c$.

Two comments are in order: First, it might appear that for the lowest order development of $\cosh(\omega t)$ one has to assume small $\omega t$, in addition to small $\theta_c$, but eq. (8) shows that these assumptions are actually equivalent. Second, Banks derives in lowest order $\dot{\theta}_c \approx \omega$ (9, p268). Using this, and additionally the small angle approximation for the arcsinh function, once again obtains (8).

In the above approximate treatment with its very simple result, the main specific effects known to occur in the domino effect, viz. anharmonicity and collective behaviour, are not present, and thus the assumptions made appear as rather strong. How accurate is this approximation then, as compared to experimental data, and to more advanced theories?

## 3  Experimental Approach[2]

A chain of dominoes is prepared on the floor (Fig. 1), with dimensions as given above (Fig. 2c). Measurements are carried out for two values of spacing ($l = 2$ cm and 3 cm), the number of dominoes is typically 20. A smartphone is placed close to the domino arrangement with the microphone in direction to the falling dominoes. An audio signal of the sounds emitted at the subsequent collisions is recorded in the smartphone (see Fig. 3) and then read out for further analysis with the application Audio Recorder[22].

From the recorded data, two variables are obtained
- time of collision between domino $n$ and $n+1$: $t_n$
- total propagation time (since the start): $T_n = \Sigma\, t_n$

---

[2] The experiments have been carried by authors LDP and EG out as student research project in the framework of a course on „Physics of Everyday Phenomena" held at the University of Geneva held by authors LD and AM.

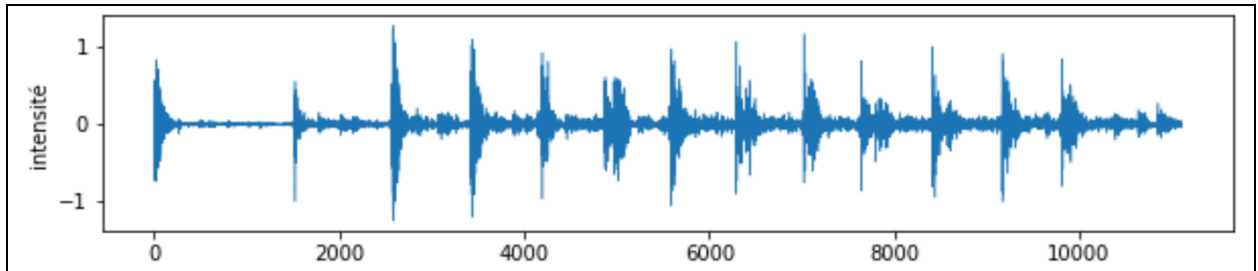

Fig. 3: Audio signal recorded of the domino chain with the microphone of a smartphone; x: axis: time ("frames"); y-axis: intensity (arbitrary units)

Of course, the collision times $t_n$ from the acoustical signal are nothing but the toppling times it takes for domino $n$ hitting the next one. Fig. 3 nicely shows how, after an initial phase, the time between subsequent collisions and thus the propagation velocity indeed become constant (2.1), i.e. we can infer an experimental value for the asymptotic toppling time as

(9) $\quad t_t \approx t_n$ for $n \gg 1$

which in practice is obtained for $n \approx 5 - 10$, see next section.

## 4  Results and comparison to theory

We first consider the case to spacing $l = 2$ cm. Fig. 4 shows the data for total time $T_n$ obtained from the acoustical measurement. Data are noisy, but a linear increase from 5 – 10 collisions on, and thus to infer the time between domino collisions from the slope of the latter. A linear fit yields the collision time for large $n$ – and by (9) thus the experimental toppling time

(10) $\quad t_t \approx (0.017 \pm 0.001)$ s

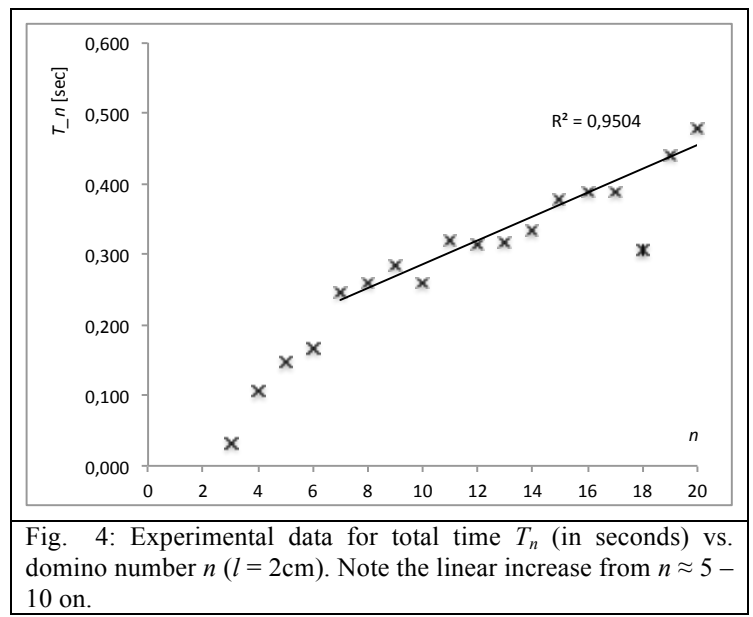

Fig. 4: Experimental data for total time $T_n$ (in seconds) vs. domino number $n$ ($l = 2$cm). Note the linear increase from $n \approx 5 - 10$ on.

(coefficient of determination $R^2 = 0.95$; one outlier excluded). The approximate result eq. (8) yields $t_t \approx 0.018$ s, in good agreement with the measurement ($s = l - d = 1.4$ cm, and $s/h = 0.34$, $\theta_c = \arcsin(s/h) = 0.35$, $\theta_c^2 \approx 0.12$; $\theta_c^3 \approx 0.04$). The propagation velocity is then $v_{as} = 118$ cm/s from experiment (111 cm/s from theory, eq. (1)).

We will now compare the values found here to the theory of van Leeuwen[7]. Using his result in its approximate form (eq. (2)), we obtain for $l = 2$ cm

(11) $\quad t_{t,\text{vLe}} \approx 0.018$ s

well consistent with the experimental result, and equal (up to the third decimal) with the approximate value given above. The corresponding velocity is 114 cm/s.

We now turn to the case of spacing $l = 3$ cm. An analysis analogous to the preceding case shows again a linear increase of $T_n$ with $n$ (even for $n$ somewhat smaller as in the previous case), and the fit yields

(12) $\quad t_t \approx (0.033 \pm 0.003)$ s.

We can compare our results to an older model by Shaw.[6] Inferring $t_t$ from the slope of his curve, one obtains $t_t = 0.033$ s consistent with eq. (12). Going beyond, we also compare the data to the prediction of the simplified model in 2.2. Equation eq. (8) yields $t_t \approx 0.031$ s, still in good agreement with the measurement, even though the small angle assumption is not well satisfied now ($s/h = 0.58$, $\theta_c = \arcsin(s/h) = 0.62$, $\theta_c^2 \approx 0.39$; $\theta_c^3 \approx 0.24$). The propagation velocity is then $v_{as} = 91$ cm/s from experiment (97 cm/s from theory).

The "frictionless" theory of van Leeuwen (eqs. (2), (3)) yields $t_{t,\text{vLe}} \approx 0.029$ and 103 cm/s, with a less good, but still satisfactory agreement with the experimental result (and with the approximate result (12) within the error of the latter). The deviation can be understood as follows: the larger the spacing, the flatter is the final position of the domino stack (see Fig. 2b), and the longer is the toppling process when dominoes slide on each other; it is thus plausible that van Leeuwen's model version neglecting friction allows for less good agreement and has to be replaced by version including friction (which is not considered here).[3]

| $v_{as}$ [cm/s] | $l$ = 2 cm | $l$ = 3cm |
|---|---|---|
| exp. | 118 | 91 |
| simplified theory (sect. 2.2) | 111 | 97 |
| van Leeuwen[7] (see eqs. (2), (3)) | 114 | 103 |
| Shaw[6] | – | 91 |

Tab. 1: Summary of experimental and theoretical values for the asymptotic velocity of the domino effect (see text for discussion).

The experimental and theoretical values for the domino experiment presented here are summarized in Tab. 1, and we now turn a brief discussion of its educational potential for physics education.

## 5  Discussion

The results in this experiment, especially the linear increase of $T_n$ with $n$ (Fig. 4) and the numerical values for the propagation speed of the domino wave show good agreement with other measurements and theory. The example can be seen as a demonstration of the use of MDETs for a student project about a phenomenon of "everyday physics", and in the same time of interest for current research: the domino effect as a paradigm of collective dynamics[5,7,8], with application e.g. in carbon nanotubes[23] or neuronal signal transduction (ref. 24, 48.3).

On the experimental level the idea of using the audio signal with a smartphone for the measurement has to be highlighted, an original idea of the student co-authors in the course "Physics of Everyday Phenomena[2]" (it later turned out that audio recording was also used in a paper on arXiv, though with a different method using more advanced equipment[25]). Indeed, smartphones come with a built-in sensor (the microphone) and recording functionality with a sufficiently good sampling quality in order to have a sufficient time resolution analysis, and the data can be easily analysed in the smartphone itself or transferred to any computer (e.g. as .csv file). In that respect, the value of MDETs compared with other methods used for the investigation of the domino effect becomes apparent, viz. photocell timing gates[6] or high speed videography[5,26], which are (much) less available and (much) more expensive. On the other hand, these methods allow for higher accuracy in the measurements, not attainable with the method presented here. But in view of the satisfactory agreement found with MDETs (see above) one may conclude that they offer a reasonable trade-off of data quality and expenditure in the context of a student research project, especially when realised autonomously and outside a lab (note that this project was carried out during the Covid university closure in 2020/21).

With these experiences we see an interesting perspective for the use of MDETs in settings like student projects, undergraduate research, etc. The case for such student research projects has been made repeatedly, emphasizing in particular its potential to foster competences one would like to develop along with content mastery, such as autonomy, curiosity, creativity, critical thinking, etc.[27,28,29] We think that the present example provides evidence for this potential and believe that MDETs in student research is a promising venue to be explored by future work, including empirical investigations on the educational outcomes.

Finally, we mention a few extensions of interest further student research projects could explore:
- testing the limits of the theoretical model, e.g. regarding the case of large domino spacings;
- testing the "no bouncing back" condition (i.e. inelasticity of the collision), for which an acoustical method appears as particularly suitable;
- investigating the influence of different surface materials;
- investigating inclined support surfaces or curved domino alignments.

On the theoretical level, a simplified model was presented here, well accessible on undergraduate level, and highlighting the central physical ideas (toppling time as determining the periodicity of the process). It yields satisfactory agreement with the measurements, and can serve as a basis for more complete models beyond the assumptions made above (sect. 2), thus completing the MDET experiment in a useful way for the purposes of physics education.

---

[3] Van Leeuwen provides numerical data for this full theory (ref. 7, Tab. II; note that the velocity data given there are in the dimensionless format $\hat{v} = v/\sqrt{gh}$). For a case similar to ours ($s/h$ = 0.6, close to our value of 0.58; $d/h$ = 0.179, about 20% larger than our value) the case without friction yields $v_{as}$ = 103 $cm/s$ (in fact equal to the value of the approximate theory), while even for a small value of the friction coefficient of $\mu$ = 0.1, he finds $v_{as}$ = 88 $cm/s$. This indeed means a considerable reduction of the asymptotic velocity by a factor of 0.85, quite close to the one we find for our experimental value compared to the frictionless case (0.88).

| https://www.youtube.com/watch?v=y97rBdSYbkg | https://youtu.be/8yYWILv91YU |

Fig. 5, left: domino chain reaction (geometric growth "in action"); right: giant version (only links given due to copyright limitations).

As a last example of the intrinsic curiosity and interest of the domino effect we mention the case of unequal, increasing sizes of the dominoes, suggested by Whitehead[30] as a nice demonstration of exponential (or geometric) growth, see Fig. 5.

# 6 References

1. SmarterEveryDay. (2017, December 11). *Dominoes - HARDCORE Mode - Smarter Every Day 182* [Video]. YouTube. https://www.youtube.com/watch?v=9hPIobthvHg&t=812s.
2. Associated Press. (2008, November 15). *Raw Video: Dutch Set Domino Record* [Video]. YouTube. https://www.youtube.com/watch?v=qD4My2Htvvw
3. Metropolis Copenhagen. (2013, July 13). *Large outdoors domino event (Copenhagen)* [Video. YouTube. https://www.youtube.com/watch?v=HMc3fvm7m7s
4. Walker, J. (2006). *The flying circus of physics*. John Wiley & Sons., french: Walker, J., & Ramonet, J. (2008). *Le cirque de la physique*. Paris: Dunod.
5. Stronge, W. J., & Shu, D. (1988). The domino effect: successive destabilization by cooperative neighbours. *Proceedings of the Royal Society of London. A. Mathematical and Physical Sciences*, *418*(1854), 155-163.
6. Shaw, D. E. (1978). Mechanics of a chain of dominoes. *American Journal of Physics*, *46*(6), 640-642.
7. Van Leeuwen, J. M. J. (2010). The domino effect. *American Journal of Physics*, *78*(7), 721-727.
8. Cantor, D., & Wojtacki, K. (2022). Effects of Friction and Spacing on the Collaborative Behavior of Domino Toppling. *Physical Review Applied*, *17*(6), 064021.
9. Banks, R. B. (2013). Towing icebergs, falling dominoes, and other adventures in applied mathematics. In *Towing Icebergs, Falling Dominoes, and Other Adventures in Applied Mathematics*. Princeton University Press.
10. McLachlan, B. G., Beaupre, G., Cox, A. B., & Gore, L. (1983). Falling dominoes (de daykin). *SIAM Review*, *25*(3), 403.
11. Bert, C. W. (1986). Falling dominoes. *SIAM Review*, *28*(2), 219-224.
12. Sun, B. H. (2020). Scaling law for the propagation speed of domino toppling. *AIP Advances*, *10*(9), 095124.
13. Song, G., Guo, X., & Sun, B. (2021). Scaling law for velocity of domino toppling motion in curved paths. *Open Physics*, *19*(1), 426-433.
14. Theron, W. F. D. (1988). The ''faster than gravity''demonstration revisited. *American Journal of Physics*, *56*(8), 736-739.
15. Härtel, H. (2000). The falling stick with a> g. *The Physics Teacher*, *38*(1), 54-55.
16. Shan, S., Shore, J. A., & Spekkens, K. (2020). The falling rod race. *The Physics Teacher*, *58*(8), 596-598.
17. Madsen, E. L. (1977). Theory of the chimney breaking while falling. *American Journal of Physics*, *45*(2), 182-184.
18. Varieschi, G., & Kamiya, K. (2003). Toy models for the falling chimney. *American Journal of Physics*, *71*(10), 1025-1031.
19. Kalmus, H. P. (1970). The inverted pendulum. *American Journal of Physics*, *38*(7), 874-878.
20. Butikov, E. I. (2001). On the dynamic stabilization of an inverted pendulum. *American Journal of Physics*, *69*(7), 755-768.
21. Chen, K. F. (2008). Standing human: an inverted pendulum. *Latin-American Journal of Physics Education*, *2*(3), 7.
22. Audio Recorder, https://f-droid.org/en/packages/com.github.axet.audiorecorder/
23. Chang, T. (2008). Dominoes in carbon nanotubes. *Physical Review Letters*, *101*(17), 175501.

24. Reece, J. B., Urry, L. A., Cain, M. L., Minorsky, P. V., & Jackson, R. B. (2013). *Campbell Biology 10th Edition* (Vol. 2). Benjamin Cummings.
25. Larham, R. (2008). Validation of a Model of the Domino Effect?. *arXiv preprint arXiv:0803.2898*.
26. Stronge, W. J. (1987). The domino effect: a wave of destabilizing collisions in a periodic array. *Proceedings of the Royal Society of London. A. Mathematical and Physical Sciences*, *409*(1836), 199-208.
27. Dolan, E. L. (2016). Course-based undergraduate research experiences: current knowledge and future directions. *Natl Res Counc Comm Pap*, *1*, 1-34. doi: 10.17226/24622
28. Ahmad, Z., & Al-Thani, N. J. (2022). Undergraduate Research Experience Models: A systematic review of the literature from 2011 to 2021. *International Journal of Educational Research*, *114*, 101996.
29. Mieg, H. A., Ambos, E., Brew, A., Galli, D., & Lehmann, J. (Eds.). (2022). *The Cambridge handbook of undergraduate research*. Cambridge University Press.
30. Whitehead, L. A. (1983). Domino``chain reaction''. *American Journal of Physics*, *51*(2), 182-182.